\def\Bbb#1{{\bf #1}}

\def\fnote#1{\footnote}

\def\cwleftpar#1#2{\leftskip #1 \rightskip #2 plus 1fill}
\def\cwrightpar#1#2{\leftskip #1 plus 1fill \rightskip #2}
\def\cwcenterpar#1#2{\leftskip #1 plus 1fill \rightskip #2 plus 1fill}
\def\cwfullpar#1#2{\leftskip#1\rightskip#2}

\def\cwoutdent#1#2{\llap{\hbox to #1{#2 \hss}}\ignorespaces}
\def\cwparbegin#1#2#3#4#5{
	\ifcase #1 \cwleftpar{#2}{#3}
	\or \cwrightpar{#2}{#3}
	\or \cwcenterpar{#2}{#3}
	\else \cwfullpar{#2}{#3}\fi
	\ifcase #4 \baselineskip = 1.5\baselineskip
	\or \baselineskip = 2\baselineskip
	\or \baselineskip = 3\baselineskip
	\else \baselineskip = 1\baselineskip\fi
	\ifdim #5 > 0in \else \noindent \fi
	\noindent\ignorespaces}
\documentclass{article}
\begin{document}
\advance \vsize by -1\baselineskip
\def\makefootline{
\ifnum\pageno = 1{\vskip \baselineskip \vskip \baselineskip }\else{\vskip \baselineskip \noindent \folio                                  \par
}\fi}

 \vspace*{2ex}
\noindent {\Huge Relative mechanical quantities\\[1ex]
		in spaces with a transport along\\[1ex] paths}\\[3ex]

\noindent Bozhidar Zakhariev Iliev
\fnote{0}{\noindent $^{\hbox{}}$Permanent address:
Laboratory of Mathematical Modeling in Physics,
Institute for Nuclear Research and \mbox{Nuclear} Energy,
Bulgarian Academy of Sciences,
Boul.\ Tzarigradsko chauss\'ee~72, 1784 Sofia, Bulgaria\\
\indent E-mail address: bozho@inrne.bas.bg\\
\indent URL: http://theo.inrne.bas.bg/$^\sim$bozho/}

\vspace*{2ex}

{\bf \noindent Published: Communication JINR, E5-94-188, Dubna, 1994}\\[1ex]
\hphantom{\bf Published: }
http://www.arXiv.org e-Print archive No.~math-ph/0310024\\[2ex]

2003 PACS numbers: 04.90.+e, 45.90.+t\\[2ex]

\noindent
{\small
The \LaTeXe\ source file of this paper was produced by converting a
ChiWriter 3.16 source file into
ChiWriter 4.0 file and then converting the latter file into a
\LaTeX\ 2.09 source file, which was manually edited for correcting numerous
errors and for improving the appearance of the text.  As a result of this
procedure, some errors in the text may exist.
}\\[1ex]

	\begin{abstract}
The concepts of relative velocity and acceleration, deviation velocity and
acceleration and relative momentum of point particles in spaces (manifolds),
the tangent bundle of which is equipped with a transport along paths, are
introduced. If the tangent bundle is endowed also with a metric, it gives
rise also to the notion of a relative energy. Certain ties between these
quantities are considered. The cases of massless particles and of special
relativity are presented in this context.
	\end{abstract}\vspace{3ex}

 {\bf 1. INTRODUCTION}
\nopagebreak

\medskip
The introduced in [1] transports along paths, which in particular can be linear [2], are applied in the present paper to defining certain mechanical quantities in spaces (manifolds), the tangent bundle of which is endowed with such a transport. Analogous problem has been considered in [3] but, in fact, in this work only linear transports along paths without self-intersections are used which is not generally necessary everywhere. We closely follow [3] without presupposing such restrictions.

All considerations in the present work are made in a (real) differentiable
manifold $M [4,5]$ whose tangent bundle $(T(M),\pi ,M)$ is endowed with a
transport along paths [1]. Here $T(M):=\cup_{x\in M}T_{x}(M), T_{x}(M)$
being the tangent to the $M$ space at $x\in M$ and $\pi :T(M)  \to M$ is such
that if $V\in T_{x}(M)$, then $\pi (V):=$x.

By $J$ and $\gamma :J  \to M$ are denoted, respectively, an arbitrary real
interval and a path in M. If $\gamma $ is of class $C^{1}$, its tangent
vector is written as $\dot\gamma$.

The transport along paths in $(T(M),\pi ,M) (cf. [1])$ is a map $I:\gamma
\to I^{\gamma }, I^{\gamma }:(s,t)\to I^{\gamma }_{s  \to t}, s,t\in J$ being
the transport along $\gamma $, where $I^{\gamma }_{s  \to t}:T_{\gamma
(s)}(M)  \to T_{\gamma (t)}(M)$, satisfy the equalities
 \[
I^{\gamma }_{t  \to r}\circ I^{\gamma }_{s  \to t}=I^{\gamma }_{s  \to
r},\quad  r,s,t\in J, \qquad (1.1)
\]
\[
I^{\gamma }_{s  \to s}={\it id}_{T_{\gamma (s)}}, \qquad  s\in J.
 \qquad (1.2)
\]
 Here ${\it id}_{X}$is the identity map of the set X.

A linear transport $(L$-transport) along paths $L$ in $(T(M),\pi ,M)$ satisfies besides (1.1) and (1.2) also the equality $(cf. [2])$

$ L^{\gamma }_{s  \to t}(\lambda U+\mu V)=\lambda L^{\gamma }_{s  \to t}U+\mu L^{\gamma }_{s  \to t}V, s,t\in J, U,V\in T_{\gamma (s)}(M). (1.3)$ In Sect. 2, the concepts of relative velocity, deviation velocity and the corresponding to them accelerations between two point particles are introduced. Sect. 3 is devoted to the relative momentum of these particles. The central role in this investigation belongs by Sect. 4. In spaces, the tangent bundle of which is endowed with a metric and a transport along paths, the relative energy of two point particles is introduced and investigated. Certain connections between the mentioned concepts are studied and the notion of a proper (rest) energy is naturally obtained. A note on the zero-mass particles case is made. Sect. 5 illustrates the considered general concepts in the case of special theory of relativity. The paper ends with some concluding remarks in Sect.6.

\medskip
\medskip
 {\bf 2. RELATIVE VELOCITY AND RELATIVE ACCELERATION}

\medskip
 Let there be given paths $x_{a}:J_{a}  \to M, a=1,2$ and $x:J  \to $M. Let
there be fixed one-to-one maps $\tau _{a}:J  \to J_{a}, a=1,2. ($The maps
$\tau _{1}$and $\tau _{2}$always exist because all real intervals are
equipollent.) Let also be given the one parameter families of paths $\{\gamma
_{s}: \gamma _{s}:J$  $  \to M, s\in J\}$ and $\{\eta _{s}: \eta _{s}:J$  $
\to M, s\in J\}$ having the properties $\gamma _{s}(r$  $):=x_{1}(\tau
_{1}(s))=:\eta _{s}(t$  $), \gamma _{s}(r$  $):=x_{2}(\tau _{2}(s))$ and
$\eta _{s}(t$  $):=x(s)$ for some $r$  ,$r$  $\in J$   and $t$  ,$t$  $\in J$
, $s\in $J.

Physically the paths $x_{1}, x_{2}$and $x$ are interpreted as trajectories (world lines) of, respectively, observed point particles 1 and 2 and a point observer observing them. The parameters $s\in J, s_{1}=\tau _{1}(s)$ and $s_{2}=\tau _{2}(s)$ are interpreted as proper times of the corresponding particles $(cf. [8]$, sect. 2).

If the particles 1 and 2 are moving along the paths $x_{1}$and $x_{2}$
respectively, then their velocities are [6,7]
\[
V_{a}:=\dot{x}_{a},\quad a=1,2.\qquad (2.1)
\]
The vectors $V_{1}$ and $V_{2}$ can not be compared as they are  defined at
different points. To compare them, we put
\[
(V_{2})_{1}
:=I^{\gamma _{s}}_{r^{\prime\prime}_s  \to r^\prime _{s}} V_{2}
\in T_{x_{1}}(M).\qquad (2.2)
\]
 As $(V_{2})_{1}$ and $V_{1}$ are defined at one and the same point, the
vector
\[
\Delta V_{21}
:=\Delta V_{21}(s;x)
:=I^{\eta _{s}}_{t^\prime_{s} \to t^{\prime\prime}_{s}}
((V_{2})_{1}-V_{1})=
\]
\[
\qquad =I^{\eta _{s}}_{t^\prime_{s}   \to t^{\prime\prime}_{s}}(I^{\gamma
_{s}}_{r^{\prime\prime}_{s}  \to r^\prime_{s}}V_{2}-V_{1})\in
T_{x(s)}(M)\qquad (2.3)
\]
is uniquely defined and represents their difference
defined at $x(s)$ with the help of I. This vector is called a {\it relative
velocity} of the second observed particle with respect to the first one (as
it is "seen" from the observer) at the point $x(s)$.

This definition of a relative velocity is a natural generalization of the Newtonian concept for a relative velocity which can be simply defined as a difference of the 3-vectors representing the velocities of the corresponding particles.

 Let the paths $\gamma _{s}, s\in J$ be of class $C^{1}$and such that the
maps
\[
 d^{\gamma }_{s}:J  \to T_{\gamma (s)}(M)=\pi ^{-1}(\gamma (s)),
 \qquad  s\in J,\qquad (2.4a)
\]
 defined by
\[
 d^{\gamma }_{s}(t):=
\int_{a}^{t}
\bigl(I^{\gamma }_{u  \to s}\gamma (u)\bigr)du,
 \qquad  s,t\in J,\qquad (2.4b)
\]
 be homeomorphisms from $J$ into $d^{\gamma }_{s}(J)$ for every $s\in J (cf.
[8]$, sect. 2).

According to [8], definition 2.3 the deviation vector of $x_{2}$
with respect to $x_{1}$relatively to $x$ at the point $x(s), s\in J$, i.e.
between the investigated particles, is
\[
h_{21}:=h_{21}(s;x)
:=\Bigl(I^{\eta _{s}}_{t^\prime_{s} \to t^{\prime\prime}_{s}}\circ
d^{\gamma _{s}}_{r^\prime _{s}} \Bigr)(r )
\]
\[
=I^{\eta _{s}}_{t_{s}^\prime   \to t^{\prime\prime}_{s}}
\int^{r_s^{\prime\prime}}_{r_s^\prime }
\bigl(I^{\gamma _{s}}_{u  \to r^\prime _{s}}\dot{\gamma}_{s}(u)
\bigr) du \in T_{x(s)}(M).\qquad (2.5)
 \]
  Let in the manifold $M$ be given also a
 covariant differentiation $\nabla $ and the deviation vector $h_{21}($of
$x_{2}$with respect to $x_{1})$ have $a C^{1}$dependence on s. Then there
arises the concept for a {\it deviation velocity} $V_{21}$between the
observed particles:  \[ V_{21}:=\frac{D}{ds}\Big|_{x}h_{21},\qquad (2.6) \]
 where $D/ds\mid_{x}:=\nabla _{\dot\gamma }$ is the covariant differentiation
along $x$ and the deviation vector is given by (2.5). This velocity has a
direct physical meaning because it can be measured. For example, if the
observer defines somehow (e.g. by radiolocation) the relative position
$h_{21}$of the observed particles, then he can find the deviation velocity
from (2.6) in which $s$ is now interpreted as observer's "proper time".

Generally speaking, the vectors $\Delta V_{21}$and $V_{21}$do not coincide
even in the Euclidean case (see [8], sect.4) in which we evidently have
\[
 \Delta V_{21}|_{E^{n}}=V_{2}-V_{1},\qquad (2.7)
\]
\[
V_{21}|_{E^{n}}=
\frac{d}{ds}
(x_{2}(\tau _{2}(s))-x_{1}(\tau _{1}(s)))
= \frac{d\tau_2(s)}{ds} \cdot V_{2}
-  \frac{d\tau_1(s)}{ds}\cdot V_{1}.  \qquad (2.8)
 \]
Nevertheless, in the Newtonian mechanics, where we have an Euclidean world
with an absolute simultaneity $(\tau _{1}=\tau _{2}=${\it id}$_{J})$, these
velocities coincide.

Let the manifold $M$ be endowed with a transport of vectors along paths and
a covariant differentiation. If $x_{1}$and $x_{2}$are $C^{2}$paths, then the
accelerations of the observed particles are
\[
  A_{a}:=
\frac{D}{ds}\Big|_{x_{a}}V_{a},\quad  a=1,2\qquad (2.9)
\]
 and we can define
in an analogous way the {\it relative acceleration} and the {\it deviation
acceleration} between them and the observer, respectively, by the equalities
\[
 \Delta A_{21}
:=I^{\eta _{s}}_{t^\prime_{s} \to t^{\prime\prime}_{s}}((A_{2})_{1}-A_{1}),
(A_{2})_{1}:=I^{\gamma _{s}}_{r^{\prime\prime}_{s}  \to
r^\prime_{s}}A_{2},\qquad (2.10)
\]
\[
A_{21}:= \frac{D}{ds}\Big|_{x}V_{21}
=\frac{D^2}{ds^2}\Big|_x h_{21}.\qquad (2.11)
\]
 The treatment of $\Delta
A_{21}$and $A_{21}$is similar to the one of $\Delta V_{21}$and $V_{21}$.

\medskip
\medskip
 {\bf 3. RELATIVE MOMENTUM}

\medskip
Let a point particle with a (rest) mass $m$ be moving along the path $\gamma
:J  \to $M. Then by definition (see $[6], ch$. III, \S3) its momentum at the
point $\gamma (s)$ is
\[
p:=p(s):=\mu (s)\dot\gamma(s),\quad s\in J
\]
,where $\mu :J  \to {\Bbb R}\backslash \{0\}$ is a scalar function with a
dimension of mass. If $m\neq 0$, then $\mu (s):=$m. If $m=0$, which is the
case, e.g., with the photons, then the momentum $p$ is considered as a
primary defined quantity and $\mu $ is obtained from the above equation. It
is important to be noted that in both the cases $\mu (s)\neq 0, s\in $J. (The
case $m=\mu (s)=0$ describes the vacuum but not a particle.)

  So, the momenta of the observed particles are
\[
 p_{a}:=p_{a}(s_{a}):=\mu _{a}(s_{a})V_{a},
\quad s_{a}=\tau _{a}(s),\ a=1,2,\ s\in J,\qquad (3.1)
\]
 where $\mu _{a}:J_{a}
\to {\Bbb R}\backslash \{0\}, a=1,2$ are scalar functions.

 As the vector
 \[
(p_{2})_{1}:=I^{\gamma _{s}}_{r^{\prime\prime}_{s}  \to
r^\prime_{s}}p_{2}\in T_{x_{1}}(M)\qquad (3.2)
\]
 is in $T_{x_{1}}(M)$, it can
be compared with $p_{1}$.  In  accordance  with this, (the vector of) the
{\it relative momentum} of the second  particle with respect to the first one
as it is "seen" from the observer  at $x(s)$ is defined by
\[
\Delta p_{21}
:=\Delta p_{21}(s;x)
:=I^{\eta _{s}}_{t^\prime_{s} \to t^{\prime\prime}_{s}}((p_{2})_{1}-p_{1})
\]
\[
\qquad
=I^{\eta _{s}}_{t^\prime_{s} \to t^{\prime\prime}_{s}}(I^{\gamma
_{s}}_{r^{\prime\prime}_{s}  \to r^\prime_{s}}p_{2}-p_{1})\in
T_{x(s)}(M).\qquad (3.3)
\]
 It is clear that in the Euclidean case (see [8], sect. 4)  the relative
momentum takes its well known Newtonian form
\[
 \Delta p_{21}|_{E^{n}}=p_{2}-p_{1}.\qquad (3.4)
\]
 If the used above
transport in $(T(M),\pi ,M)$ is linear (see Sect. $1, eq. (1.3)$ or [2]),
then due to $(3.1)-(3.3)$ and $(2.2)-(2.3)$ the following equalities are
valid
\[
 (p_{2})_{1}=\mu _{2}(s_{2})(V_{2})_{1},\qquad (3.5)
\]
\[
 \Delta p_{21}=\mu
_{2}(s_{2})\Delta V_{21}+[\mu _{2}(s_{2})/\mu _{1}(s_{1})-1]I^{\eta
_{s}}_{t^\prime_{s}   \to t^{\prime\prime}_{s}}p_{1}.\qquad (3.6)
\]

\medskip
 {\bf 4. RELATIVE ENERGY}

\medskip
Let in the tangent bundle $(T(M),\pi ,M)$ be given a transport along
paths I and a real bundle metric $g$, i.e., $[5] a$ map $g:x\to g_{x}, x\in M$, where the maps $g_{x}:T_{x}(M)\otimes T_{x}(M)  \to {\Bbb R}$ are bilinear, nondegenerate and symmetric. For brevity, the defined by $g$ scalar products of $X,Y\in T_{y}(M), y\in M$ will be denoted by a dot $(\cdot )$, i.e. $X\cdot Y:=g_{y}(X,Y)$. The scalar square of $X$ will be written as $(X)^{2}$for it has to be distinguished from the second component $X^{2}$of $X$ in some local basis (in the case when $\dim(M)>1)$. As $g$ is not supposed to be positively defined, $(X)^{2}$can take any real values.

By definition the {\it relative energy} of the second particle with respect
to the first one is called the (scalar) quantity
\[
 E_{21}:=E_{21}(s):=\epsilon ((V_{1}(s_{1}))^{2})p_{21}\cdot V_{1}(s_{1})
\]
\[
=\epsilon ((V_{1}(s_{1}))^{2})\bigl(I^{\gamma
_{s}}_{r^{\prime\prime}_{s}  \to r^\prime_{s}}p_{2}(s_{2})\bigr)\cdot
V_{1}(s_{1}),\qquad (4.1)
\]
 where $\epsilon (\lambda ):=-1$ for $\lambda <0$
and $\epsilon (\lambda ):=+1$ for $\lambda \ge 0$. The introduction of the
multiplier $\epsilon $ is due to the fact that if the particles coincide,
i.e., if we apply (4.1) to one and the same particle, then the so obtained
quantity has a meaning of a proper energy of that particle (see below) and
according to the accepted opinion [6,7] it must be positive.

 If there exists $s_{0}\in J$ such that $x_{1}(\tau _{1}(s_{0}))=x_{2}(\tau
_{2}(s_{0}))$, i.e. if at the "moment$" s=s_{0}$the trajectories of the
observed particles intersect each other, then from (4.1) and (1.2) we get
\[
 E_{21}(s_{0})=\epsilon ((V_{1}(\tau _{1}(s_{0})))^{2})p_{2}(\tau
_{2}(s_{0}))\cdot V_{1}(\tau _{1}(s_{0})).\qquad (4.2)
\]
 In the case of the
space-time of general relativity, this expression coincides with the given in
$[6], ch$. III, $\S6, eq. (23)$ definition for a relative energy which has
the "bad" property that it is valid only for the "moment$" s=s_{0}$. So it
does not allow the evolution of the relative energy in time to be studied.
Evidently, our definition (4.1) is free from this deficiency.

 Analogously to (4.1), the relative energy of the first particle with respect
to the second one is
 \[
 E_{12}:=E_{12}(s):=\epsilon ((V_{2}(s_{2}))^{2})p_{12}\cdot V_{2}(s_{2})
\]
\[
=\epsilon ((V_{2}(s_{2}))^{2})\bigl(I^{\gamma _{s}}_{r_{s}^\prime   \to
r^{\prime\prime}_{s}}p_{1}(s_{1})\bigr)\cdot V_{2}(s_{2}).\qquad (4.3)
\]

If we use arbitrary transports along paths, then, generally, the quantities
$E_{12}, E_{21}, \Delta p_{21}$and $\Delta p_{12}$are not connected somehow
with each other. From the view point of the existence of a certain connection
between them an essential role is played by the transports along paths which
are consistent (at least along the paths $\gamma _{s}$and $\eta _{s}, s\in
J)$ with the fibred metric $g$, i.e. for which $(cf. [9,10])$
\[
I^{\gamma }_{s  \to t}(U\cdot V)=(I^{\gamma }_{s  \to t}U)\cdot
(I^{\gamma }_{s  \to t}V)\qquad (4.4)
\]
 for arbitrary $\gamma :J  \to M,
s,t\in J$ and $U,V\in T_{\gamma (s)}(M). ($For further considerations it is
enough that this equality be valid only for $\gamma \in \{\gamma _{s}, \eta
_{s}: s\in J\}.)$

 If I and $g$ are consistent, then the relative momentum (3.3) and the
relative energy (4.1), as one can easily prove, are connected by the relation
 \[
 E_{21}(s)=\epsilon ((V_{1}(s_{1}))^{2})[\Delta p_{21}\cdot I^{\eta
_{s}}_{t_{s}^\prime   \to t^{\prime\prime}_{s}}V_{1}(s_{1})+p_{1}\cdot
V_{1}(s_{1})].\qquad (4.5)
\]

 If the transport along paths I is consistent with
the operation multiplication with real numbers (see [10], example 3.2), i.e.
\[
 I^{\gamma }_{s  \to t}(\lambda U)=\lambda I^{\gamma }_{s  \to t}U,
\quad \lambda \in {\Bbb R},\ U\in T_{\gamma (s)}(M),\qquad (4.6)
\]
 then (3.5) holds and after
its substitution into (4.1), one gets
\[
 E_{21}=\epsilon ((V_{1}(s_{1}))^{2})\mu _{2}(s_{2})(V_{2})_{1}\cdot
V_{1}(s_{1}).\qquad (4.7)
\]
If the equalities (4.4) and (4.6) are
simultaneously valid, then with a direct verification we confine ourselves to
that the relative energies $E_{21}$and $E_{12}$are connected by
\[
 \epsilon ((V_{2}(\tau _{2}(s)))^{2})\mu _{1}(\tau _{1}(s))E_{21}(s)=\epsilon
((V_{1}(\tau _{1}(s)))^{2})\mu _{2}(\tau _{2}(s))E_{12}(s).  \qquad (4.8)
\]

In  particular, this equality is  true  for  every  $L$-transport consistent
with the metric.

Let us apply definition (4.1) only to the first observed particle, for which
it is enough to put in it $r$  $=r$  , $x_{2}=x_{1}$and $\tau _{2}=\tau
_{1}$, or equivalently to replace the subscript 2 with 1. Using (1.2), we see
that the energy of this particle (with respect to itself) is
\[
  E_{11}(s)=\epsilon ((V_{1}(s_{1}))^{2})p_{1}(s_{1})\cdot V_{1}(s_{1})=\mu
_{1}(s_{1})\mid (V_{1}(s_{1}))^{2}\mid
\]
\[
 =\mid (p_{1}(s_{1}))^{2}\mid /\mu _{1}(s_{1}),\qquad (4.9)
\]
 where $\mid \lambda \mid :=\epsilon (\lambda )\lambda $ is the absolute
value of $\lambda \in {\Bbb R}$.

The quantity $E_{11}$may be called a {\it proper} (or rest) {\it energy} of the considered particle. If $m_{1}>0$, then $\mu _{1}(s_{1})=m_{1}$and consequently $E_{11}\ge 0$. If $m_{1}(V_{1}(s_{1}))^{2}\neq 0$, then $E_{11}>0$ which corresponds to the most popular case of massive material particle.

If $m_{1}\neq 0$, then $\mu _{1}(s_{1}):=m_{1}$and due to (4.9) the proper
energy $E_{11}$is proportional to $m_{1}$, so $E_{11}$is $a C^{\infty
}$function of $m_{1}$for $m_{1}\in {\Bbb R}\backslash \{0\}$. From here comes
the mind on $E_{11}$to be imposed the additional restriction for continuous
dependence of $m_{1}$at the point $m_{1}=0$, i.e. one may want
\[
 E_{11}=0\quad for\ m_{1}=0,\qquad (4.10)
\]
 or, equivalently,
\[
\lim_{m_i\to 0}E_{11}=0,\qquad (4.11)
\]
which has far going physical
corollaries. In fact, (4.9) shows the equivalence of (4.11) with
\[
 (V_{1}(s_{1}))^{2}=0\quad for\ m_{1}=0,\qquad (4.12)
\]
 or, which is all the same, with
\[
 (p_{1}(s_{1}))^{2}=0\quad  for\ m_{1}=0.\qquad (4.13)
\]
 These relations are
a direct generalization of the well known fact from the special and general
relativity that the massless particles are moving with the velocity of light,
i.e. that their world lines lie on the light cone described by (4.12).

We want to note that without further assumptions $(V_{1}(s_{1}))^{2}=0$ does not imply $m_{1}=0$.

The energies $E_{21}($or $E_{12})$ and $E_{11}$may be connected with the components of $\Delta p_{21}($or $\Delta p_{12}), p_{21}($or $p_{12})$ and $p_{1}$in some local bases in the following way.

Let $(V_{1})^{2}\neq 0$. Along $x_{1}$we define a field of basis
$\{\lambda_{i}\}$, i.e. the vectors $\lambda _{i}\mid _{\gamma (s)}\in
T_{\gamma (s)}(M)$ form a basis in $T_{\gamma (s)}(M)$, such that $\lambda
_{1}:=V_{1}\cdot \mid (V_{1})^{2}\mid ^{-1/2}$and $\lambda _{1}\cdot \lambda
_{i}=0$ for $i\neq 1 ($if $\dim(M)>1)$. Here and henceforth the Latin indices
run from 1 to $\dim(M). ($In this case the concrete choice of $\lambda
_{i}$for $i\neq 1$ is insignificant.) So $(\lambda _{1})^{2}=\epsilon
((V_{1})^{2})$, due to which the component $A^{1}$of any vector field
$A=A^{i}\lambda _{i}$along $x_{1}$in $\{\lambda _{i}\}$ is
\[
A^{1}=A\cdot \lambda _{1}/(\lambda _{1})^{2}=\epsilon ((V_{1})^{2})(A\cdot
V_{1})\mid (V_{1})^{2}\mid ^{-1/2}.\qquad (4.14)
\]

Applying this equality to
$p_{1}, p_{21}$and $\Delta \pi _{21}:=\Delta p_{21}$
$_{x=x_{1}}=(p_{2})_{1}- -p_{1}$, the last vector being the relative momentum
of the second particle with respect to the first one as it is "seen" from the
latter, and using (4.9) and (4.1), we find:
\[
 p^{1}_{1}=E_{11}\mid (V_{1})^{2}\mid ^{-1/2}, p^{i}_{1}=0
 \quad  for\ i\neq 1,\qquad (4.15)
 \]
 \[
  (p_{2})^{1}_{1}=E_{21}\mid (V_{1})^{2}\mid ^{-1/2},\qquad
  \qquad (4.16)
 \]
 \[
 \Delta \pi^{i}_{21}=(p_{2})^{1}_{1}-p^{1}_{1}=(E_{21}-E_{11})\mid
 (V_{1})^{2}\mid ^{-1/2}.\qquad (4.17)
 \]

  Let $(V_{1})^{2}\neq 0$ and $g$ and I
 be consistent, i.e. (4.4) be valid. Defining along $x a$ basis $\{l_{i}\}$
 such that $l_{1}:=I^{\eta _{s}}_{t_{s}^\prime   \to
 t^{\prime\prime}_{s}}\lambda _{1}$and $l_{1}\cdot l_{i}=0$ for $i\neq 1$, we
 see that the first component of $\Delta p_{21}$in $\{l_{i}\}$ is
\[
 \Delta p^{1}_{21}=\Delta p_{21}\cdot l_{1}/(l_{1})^{2}=\Delta \pi _{21}\cdot
 \lambda _{1}/(\lambda _{1})^{2}=\Delta \pi ^{1}_{21}
\]
\[
=(E_{21}-E_{11})\mid (V_{1})^{2}\mid ^{-1/2}.   \qquad (4.18)
 \]

  If $(V_{1})^{2}=0$, then (see $(4.9))
 E_{11}=0$ and the invariant $(p_{2})_{1}\cdot V_{1}= =\Delta \pi _{21}\cdot
 V_{1}=E_{21}$cannot be connected with some component of $(p_{2})_{1}($or
 $\Delta \pi _{21})$ in a given local basis. In this case, we can say that
 the relative energy $E_{21}$is spread over all the components of
 $(p_{2})_{1}($or $\Delta \pi _{21})$ and with basis transformations it
 cannot be connected with a single component of that vector.

 If $(V_{2})^{2}\neq 0$ and I and $g$ are consistent, i.e. (4.4) holds, then
 defining along $x_{1}a$ basis $\{\lambda _{i^\prime }\}$ such that $\lambda
 _{1^\prime }:= :=I^{\gamma _{s}}_{r_{s}^{\prime\prime}  \to r^\prime
 _{s}}(V_{2}\cdot \mid (V_{2})^{2}\mid ^{-1/2})$ and $\lambda _{1^\prime
 }\cdot \lambda _{i^\prime }=0$ for $i\neq 1$, we find the first component of
 $p_{1}$in $\{\lambda _{i}\}$ as
 \[
   p^{1^\prime }_{1}=p_{1}\cdot \lambda _{1^\prime }/(\lambda
 _{1})^{2}=p_{1}\cdot \bigl(I^{\gamma _{s}}_{r_{s}^{\prime\prime}  \to
 r^\prime _{s}}(V_{2}\cdot \mid (V_{2})^{2}\mid ^{-1/2})\bigr)\cdot
 \epsilon ((V_{2})^{2})
 \]
 \[
 =\epsilon ((V_{2})^{2})(I^{\gamma _{s}}_{r_{s}^\prime   \to
 r^{\prime\prime}_{s}}p_{1})\cdot V_{2}\cdot \mid (V_{2})^{2}\mid
 ^{-1/2}=E_{21}\mid (V_{2})^{2}\mid ^{-1/2}.
  \qquad (4.19)
 \]

 At the end of this   section we want to stress the fact that  all relative
 energies $E_{11}, E_{21}, E_{12}$and $E_{22}$, connected with the  consi
 dered observed particles, are not arbitrary, as they are  connected with the
 invariant $(\Delta \pi _{21})^{2}$by
\[
   (\Delta \pi _{21})^{2}=\epsilon ((V_{1})^{2})\mu _{1}E_{11}+\epsilon
 ((V_{2})^{2})\mu _{2}E_{22}-2\epsilon ((V_{1})^{2})\mu _{1}E_{21}.\qquad
 (4.20)
 \]
  This follows from $(\Delta \pi
 _{21})^{2}=((p_{2})_{1})^{2}-2(p_{2})_{1}\cdot p_{1}+(p_{1})^{2}$and the
 definitions of the corresponding energies. If the transport and the metric
 are consistent, then this equality can be written in a more symmetric form
 as
 \[
   (\Delta p_{21})^{2}=(\Delta \pi _{21})^{2}=\epsilon ((V_{1})^{2})\mu
 _{1}E_{11}+\epsilon ((V_{2})^{2})\mu _{2}E_{22}
 \]
 \[
 -\epsilon ((V_{1})^{2})\mu _{1}E_{21}-\epsilon ((V_{2})^{2})\mu
 _{2}E_{12},\qquad (4.21)
 \]
  where we have used (4.8).

\medskip
\medskip
 {\bf 5. EXAMPLE: SPECIAL RELATIVITY}

\medskip
The purpose of this section is to find explicit forms of the introduced relative quantities in the concrete case of special relativity. (As a standard reference to the problems of this theory see, e.g., [6,7].)

Let there be given a standard (4-dimensional, flat, with signature (+---)) Minkowski's space-time $M^{4}$, in which as a concrete realization of the general transport along paths the parallel transport along them will be used. Let two point particles 1 and 2 with masses $m_{1}\neq 0$ and $m_{2}\neq 0$ be moving in $M^{4}$with constant 3-velocities ${\bf v}_{1}$and ${\bf v}_{2}$, respectively, with respect to a given frame of reference. Then, their world lines are $x_{a}(s_{a})=(ct,t{\bf v}_{a})+y_{a}, a=1,2$, where $c$ is the velocity of light in vacuum, $t$ is the time in the used frame, $s_{a}:=\tau _{a}(t):=t(1-{\bf v}^{2}_{a}/c^{2})^{1/2}, a=1,2$ are the corresponding proper times and $y_{1},y_{2}\in M^{4}$are fixed.

  According to (4.1), the 4-velocities [6] of the particles are
 \[
   V_{a}=(c,{\bf v}_{a})(1-{\bf v}^{2}_{a}/c^{2})^{-1/2}, a=1,2\qquad (5.1)
 \]
 and hence
 \[
   (V_{a})^{2}=(c^{2}-{\bf v}^{2}_{a})\bigl( (1-{\bf
 v}^{2}_{a}/c^{2})^{-1/2}\bigr)^{2}=c^{2}, \epsilon ((V_{a})^{2})=+1,
 a=1,2.  \qquad (5.2)
 \]
 Due to this by using $(3.1)-(3.3), (4.1)$ and (4.9), we get:
 \[
   p_{a}=m_{a}(c,{\bf v}_{a})(1-{\bf v}^{2}_{a}/c^{2})^{-1/2}, a=1,2 (\mu
 _{1}=m_{1}, \mu _{2}=m_{2}),\qquad (5.3a)
 \]
 \[
 (p_{1})_{2}=p_{1},
 (p_{2})_{1}=p_{2}, \Delta p_{21}=\Delta \pi _{21}=p_{2}-p_{1},\qquad (5.3b)
 \]
\[
 E_{21}=m_{2}c^{2}(1-{\bf v}_{1}\cdot {\bf v}_{2}/c^{2})[(1-{\bf
 v}^{2}_{1}/c^{2})(1-{\bf v}^{2}_{2}/c^{2})]^{-1/2},\qquad (5.4a)
 \]
 \[
 E_{12}=m_{1}c^{2}(1-{\bf v}_{1}\cdot {\bf v}_{2}/c^{2})[(1-{\bf
 v}^{2}_{1}/c^{2})(1-{\bf v}^{2}_{2}/c^{2})]^{-1/2},\qquad (5.4b)
 \]
 \[
  E_{11}=m_{1}c^{2}, \qquad  E_{22}=m_{2}c^{2}.\qquad (5.4c)
 \]

  Evidently, $E_{11}$ and  $E_{22}$are the proper (rest) energies of the
 particles. If, e.g., ${\bf v}_{1}=0$, then $E_{21}=m_{2}c^{2}(1-{\bf
 v}^{2}_{2}/c^{2})^{-1/2}=E_{2}$is the energy of the second particle with
 respect to the used frame [6].

 If $m_{1}\neq 0$ and $m_{2}=0$, then in the above considerations one has to
 replace $x_{2}$and $s_{2}$, respectively, by $x_{2}(s_{2})=(ct,ct{\bf
 n}_{2})+y_{2}$and $s_{2}=t$, where ${\bf n}_{2}$is a unit 3-vector $({\bf
 n}^{2}_{2}=1)$ showing the direction of movement of the second particle,
 i.e. ${\bf v}_{2}=c{\bf n}_{2}$, and, consequently
 \[
   V_{2}=c(1,{\bf n}_{2}), (V_{2})^{2}=0, \epsilon ((V_{2})^{2})=+1.\qquad
 (5.5)
 \]
  If $E_{2}$is the energy of the second particle with respect to the
 given frame, then its 4-momentum is [6]
 \[
  p_{2}=(E_{2}/c,{\bf p}_{2})=(E_{2}/c,(E_{2}/c){\bf n}_{2})=(E_{2}/c)(1,{\bf
 n}_{2})=(E_{2}/c^{2})V_{2}\qquad (5.6)
 \]
  and due to (3.1), we have
 \[
 \mu _{2}=\mu _{2}(s_{2})=E_{2}/c^{2}.\qquad (5.7)
 \]
  In this case, (5.3b) is  also true and (5.4) take the form:
 \[
   E_{21}=E_{2}(1-{\bf v}_{1}\cdot {\bf n}_{2}/c)(1-{\bf
 v}^{2}_{1}/c^{2})^{-1/2},\qquad (5.8a)
 \]
 \[
  E_{12}=m_{1}c^{2}(1-{\bf v}_{1}\cdot {\bf n}_{2}/c)(1-{\bf
 v}^{2}_{1}/c^{2})^{-1/2},\qquad (5.8b)
 \]
 \[
 E_{11}=m_{1}c^{2}, \qquad  E_{22}=0,\qquad (5.8c)
 \]
  the last of which is in accordance with (4.10).

Evidently, $E_{21}=E_{2}$for ${\bf v}_{1}=0$, due to which (5.8a) expresses
 the usual Doppler effect in terms of energies of the corresponding particles
 [6]. In fact, if we have a moving with $a 3$-velocity ${\bf v}_{1}={\bf v}$
 source of massless particles (e.g. photons) with 3-velocities ${\bf
 v}_{2}=c{\bf n}$ and energy (with respect to the source$) E_{21}=E_{0}$,
 which are registered by immovable in this frame  observer, we will find that
 the particles are with energy $E=E_{2}$, which due to (5.8a) is
 \[
  E=E_{0}(1-{\bf v}\cdot {\bf n}/c)^{-1}(1-{\bf v}^{2}/c^{2})^{1/2}.\qquad
 (5.9)
 \]

  The corresponding formulae for $m_{1}=0$ and $m_{2}\neq 0$ are
 obtained from the above ones by means of the change $1  \rightarrow 2
 \rightarrow 1$ of the subscripts in them.

In the case when $m_{1}=m_{2}=0$, we have $x_{a}(s_{a})=(ct,ct{\bf
 n}_{a})+y_{a}, s_{a}=t, a=1,2$, so
\[
 {\bf v}_{a}=c{\bf n}_{a}, {\bf n}^{2}_{a}=1, V_{a}=c(1,{\bf n}_{a}),
 (V_{a})^{2}=0, \epsilon ((V_{a})^{2})=+1,\qquad (5.10)
 \]
 \[
  p_{a}=(E/c)(1,{\bf n}_{a}), \mu _{a}=\mu _{a}(t)=E_{a}/c^{2},
  \quad a=1,2,\qquad  (5.11)
 \]
  and the equations (5.3b) remain the same. Hence:
 \[
   E_{21}=E_{2}(1-{\bf n}_{1}\cdot {\bf n}_{2}), E_{12}=E_{1}(1-{\bf
 n}_{1}\cdot {\bf n}_{2}),\qquad (5.12a)
 \]
 \[
 E_{11}=E_{22}=0.\qquad (5.12b)
 \]

 So, if ${\bf n}_{1}={\bf n}_{2}$, then $E_{21}=E_{12}=0$ and vice versa.

At the end, we shall consider the concepts of relative velocity and
 deviation velocity in special relativity.

Let $K$ be a fixed inertial frame of reference in which an arbitrary moving
 particle 2 has $a 4$-radius-vector $x_{2}(s_{2})\mid _{K}=(ct,{\bf
 x}_{2}(t)), s_{2}=t(1-{\bf v}^{2}_{2}/c^{2})^{1/2}, {\bf v}_{2}=$  $_{2}$,
 where $t$ is the time in K. Let the inertial frame $K^\prime $ be attached
 to the particle 1 having in $K a$ world line $x_{1}(s_{1})\mid
 _{K}=(ct,t{\bf v}_{1}), {\bf v}_{1}=$const, $s_{1}=t(1-{\bf
 v}^{2}_{1}/c^{2})^{1/2}$. The world line of the observer is completely
 arbitrary.

 In the frame $K$, we have
 \[
   V_{a}|_{x}
 =\frac{d x_a|_K}{ds_a} = (c,{\bf v}_{a})(1-{\bf v}^{2}_{a}/c^2)^{-1/2},
 \quad a=1,2,\qquad (5.13a)
 \]
  and in $K^\prime $,  we get
 \[
 V_{1}|_{K^\prime }=(c,{\bf 0}),
 \quad V_{2}|_{K^\prime }=(c,{\bf v}')
 (1-{\bf v'_2}^2/c^2)^{-1/2},\qquad (5.13b)
 \]
  where ${\bf v}$   is the  3-velocity of the particle 2 in $K^\prime  ($i.e. with respect to the
 particle 1) in a sense of special relativity (see [6]).

Consequently, as we are working in a pseudo-Euclidean case, due to (2.6) the
 relative velocity is $\Delta V_{21}=V_{2}-V_{1}$. So, we get:
\[
  \Delta V_{21}|_{K}
 =(1-{\bf v}^{2}_{2}/c^{2})^{-1/2}(c,{\bf v}_{2})
 - (1-{\bf v}^{2}_{1}/c^{2})^{-1/2}(c,{\bf v}_{1}),\qquad (5.14a)
 \]
 \[
 \Delta V_{21}|_{K^\prime }
 =(1-{\bf v'_2}^{2}/c^{2})^{-1/2}(c,{\bf v})
 -(c,{\bf 0}).\qquad (5.14b)
 \]
 Besides, in the pseudo-Euclidean case  $h_{21}=x_{2}(s_{2})-x_{1}(s_{1})$,
 so that:
 \[
  h_{21}|_{K}=(0,{\bf x}_{2}(t)-t{\bf v}_{1}),\qquad (5.15a)
 \]
 \[
 h_{21}|_{K^\prime }=(0,{\bf x} ),  \qquad  (5.15b)
 \]
  where ${\bf x}$   is obtained from ${\bf x}_{2}(t)$ by a Lorentz
 transformation describing the transition from $K$ to $K^\prime  [6]$.

 Due to (2.6) the deviation velocity is
 \[
  V_{21}=dh_{21}/ds=(ds_{1}/ds)(dh_{21}/ds_{1})=(dt/ds)(dh_{21}/dt),
 \]
  where $s_{1}=t^\prime $ is the time in $K^\prime  [6]$, from where, we get:
 \[
 V_{21}|_{K}
 = \frac{dt}{ds}(0,{\bf v}_{2}-{\bf v}_{1}),\qquad (5.16a)
 \]
 \[
 V_{21}|_{K^\prime }=\frac{ds_1}{dt} (0,{\bf v}  ).   \qquad (5.16b)
 \]

  So, if the observer coincides with the first particle, then $s_{1}=s$ and
 $V_{21}$ $_{K^\prime }=(0,{\bf v}$  ). This shows that in fact the deviation
 velocity is a direct generalization of the relative velocity in a sense of
 special relativity.

\medskip
\medskip
 {\bf 6. COMMENTS}

\medskip
In this work many times was met the problem for comparing (defining the difference of) two defined at different points vectors. Below is presented a general scheme for the used in the present work method in a manifold $M$ endowed with a transport along paths I in its tangent bundle.

Let $A_{a}\in T_{z_{a}}(M), a=1,2, \gamma :J  \to M$ and $\gamma
 (s_{a})=z_{a}, a=1,2$  for some $s_{1},s_{2}\in J$ and $z_{1},z_{2}\in $M.
 Let
\[
  (A_{2})_{1}:=I^{\gamma }_{s_{2}}A_{2}, (A_{1})_{2}:=I^{\gamma
 }_{s_{1}}A_{1}.\qquad (6.1)
 \]
  Now instead of $A_{1}$and $A_{2}$one can
 compare the vectors $A_{1}$and $(A_{2})_{1}$, or equivalently the vectors
 $A_{2}$and $(A_{1})_{2}$. The corresponding differences (defined by I), by
 definition, are
 \[
  \Delta A_{21}:=(A_{2})_{1}-A_{1}, \Delta A_{12}:=(A_{1})_{2}-A_{2}.\qquad
 (6.2)
 \]
  Evidently, for a linear transport along paths $L$ these two
 quantities are connected by
\[
   \Delta A_{12}
=L^{\gamma }_{s_{1}\to s_2}\Delta A_{21},
 \qquad
 \Delta A_{21}=L^{\gamma }_{s_{2}\to s_1}\Delta A_{12}.\qquad (6.3)
 \]

I n manifolds with a transport of vectors along paths and a covariant
 differentiation there arises a "mixed" acceleration
$\frac{D}{ds}\Big|_{x}\Delta V_{21}$, but there are not physical reasons
that it plays some significant role.

Let $X_{a}$denote one of the vector fields of velocity, acceleration or
 momentum of the a-th, $a=1,2$ particle. In sections 2 and 3 we introduce the
 quantities
 \[
 (X_{2})_{1}:=I^{\gamma _{s}}_{r_{s}^{\prime\prime}  \to
 r^\prime_{s}}X_{2}\in T_{x_{1}}(M),\qquad (6.4)
 \]
 \[
 \Delta X_{21}:=\Delta  X_{21}(s;x):=I^{\eta _{s}}_{t_{s}^\prime   \to
 t^{\prime\prime}_{s}}((X_{2})_{1}-X_{1})
 \]
 \[
  =I^{\eta _{s}}_{t_{s}^\prime   \to t^{\prime\prime}_{s}}(I^{\gamma
 _{s}}_{r_{s}^{\prime\prime}  \to r^\prime_{s}}X_{2}-X_{1})\in
 T_{x(s)}(M).\qquad (6.5)
 \]

  Analogously, if $\eta ^{*}_{s}:J^{*}  \to M, s\in
 J, \eta ^{*}_{s}(t^{*}_{s}):=x_{2}(s_{2})$ and $\eta ^{*}_{s}(t^{**}_{s}):=
 :=x(s)$ for some $t^{*}_{s},t^{**}_{s}\in J^{*}_{s}$and using the same paths
 $\gamma _{s}, s\in J$, we can define the quantities
 \[
  (X_{1})_{2}:=I^{\gamma _{s}}_{r_{s}^\prime   \to
 r^{\prime\prime}_{s}}X_{1}\in T_{x_{2}}(M),\qquad (6.6)
 \]
 \[
 \Delta  X_{12}:=\Delta X_{12}(s;x):=I^{\eta ^{*}_{s}}_{t^{*_{  \to
 t}}_{s}}((X_{1})_{2}-X_{2})
 \]
 \[
 =I^{\eta ^{*}_{s}}_{t^{*_{  \to t}}_{s}}(I^{\gamma _{s}}_{r_{s}^\prime
 \to r^{\prime\prime}_{s}}X_{1}-X_{2})\in T_{x(s)}(M),\qquad (6.7)
 \]
 the  latter of which, in a case of linear transport along paths $L$, due to
 (1.1) and (1.3) is connected with $\Delta X_{21}$by
 \[
  \Delta X_{12}=-L^{\eta ^{*}_{s}}_{t^{*_{  \to t}}_{s}}\circ L^{\gamma
 _{s}}_{r_{s}^{\prime\prime}  \to r^\prime_{s}}\circ I^{\eta
 _{s}}_{t_{s}^{\prime\prime}  \to t^\prime_{s}}(\Delta X_{21}).\qquad (6.8)
 \]

   If $L^{\gamma }$does not depend on $\gamma $ or if $\eta ^{*}_{s}$ is a
 product of $\gamma _{s}$and $\eta _{s}$and the equalities (2.6) and (2.7) of
 [1] are true, then according to [1], proposition 3.4 the last equality
 reduces to
\[
  \Delta X_{12}=-\Delta X_{21}.\qquad (6.9)
\]

 For some purposes, in (4.1) one
 can put $\epsilon (0)=-1$ instead of $\epsilon (0)=+1$. Our general results
do not depend on that choice.

 Opposite to (4.11), if we admit that
$\lim_{m_2\to 0}E_{21}=0$, i.e. a
continuous dependence of $E_{21}$on $m_{2}$, then we arrive at an explicit
contradiction with the physical reality. Namely, if this is so, then due to
$\mu _{2}(s_{2})\neq 0$ from (4.1), we get $V_{2}(s_{2})=0$ for $m_{2}=0$,
which contradicts the fact that we are dealing with a material particle but
not with the vacuum. Besides, the equality $E_{21}=0$ for $m_{2}=0$ means
that any massless particle, e.g. a photon, has zero (relative) energy with
respect to any other particle, something which, evidently, is not true.

\medskip
\medskip
 {\bf ACKNOWLEDGEMENT}

\medskip
This research was partially supported by the Fund for Scientific Research of Bulgaria under contract Grant No. $F 103$.

\medskip
\medskip
 {\bf REFERENCES}

\medskip
1.  Iliev B.Z., Transports along paths in fibre bundles. General theory, JINR Communication $E5-93-299$, Dubna, 1993.\par
2.  Iliev B.Z., Linear transports along paths in vector bundles. I. General theory, JINR Communication $E5-93-239$, Dubna, 1993.\par
3.  Iliev B.Z., Relative velocity, momentum and energy of point particles in spaces with general linear transport, JINR Communication $E2-89-616$, Dubna, 1989.\par
4.  Kobayashi S., K. Nomizu, Foundations of differential geometry, vol.1, Interscience publishers, New-York-London, 1963.\par
5.  Dubrovin B.A., S.P. Novikov, A.T. Fomenko, Modern geometry,
Nauka, Moscow, 1979 (in Russian).\par
6.  Synge J.L., Relativity: The general theory, North-Holland Publ. Co., Amsterdam, 1960.\par
7.  Hawking S.W., G.F.R. Ellis, The large scale structure of space-time, Cambridge Univ. Press, Cambridge, 1973.\par
8.  Iliev B.Z., Deviation equations in spaces with a transport along paths, JINR Communication $E2-94-40$, Dubna, 1994.\par
9.  Iliev B.Z., Linear transports along paths in vector bundles. IV. Consistency with bundle metrics, JINR Communication $E5-94-17$, Dubna, 1993.\par
10.  Iliev B.Z., Transports along paths in fibre bundles III. Consistency with bundle morphisms, JINR Communication $E5-94-41$, Dubna, 1994.

\newpage
\medskip
\medskip
\noindent Iliev B. Z.\\[5ex]

\noindent Relative Mechanical Quantities\\
 in Spaces with a Transport along Paths\\[5ex]

\medskip
\medskip
\medskip
The concepts of relative velocity and acceleration, deviation velocity and
acceleration and relative momentum of point particles in spaces (manifolds),
the tangent bundle of which is equipped with a transport along paths, are
introduced. If the tangent bundle is endowed also with a metric, it gives
rise also to the notion of a relative energy. Certain ties between these
quantities are considered. The cases of massless particles and of special
relativity are presented in this context.\\[5ex]

\medskip
\medskip
The investigation has been performed at the Bogoliubov Laboratory of
Theoretical Physics, JINR.

\end{document}